\documentclass{aa}
\usepackage[varg]{txfonts}
\usepackage{longtable}
\usepackage{lscape}
\usepackage{graphicx}
\newcommand{\kmprs}  {\mbox{\rm km\,s$^{-1}$}}
\newcommand{\feh} {\mbox{\rm [Fe/H]}}
\newcommand{\fehI} {\mbox{\rm [Fe/H]$_{\rm I}$}}
\newcommand{\fehII} {\mbox{\rm [Fe/H]$_{\rm II}$}}

\newcommand{\xh} {\mbox{\rm [X/H]}}
\newcommand{\ch} {\mbox{\rm [C/H]}}

\newcommand{\oh} {\mbox{\rm [O/H]}}

\newcommand{\nah} {\mbox{\rm [Na/H]}}

\newcommand{\mgh} {\mbox{\rm [Mg/H]}}

\newcommand{\alh} {\mbox{\rm [Al/H]}}
\newcommand{\sih} {\mbox{\rm [Si/H]}}
\newcommand{\sh} {\mbox{\rm [S/H]}}
\newcommand{\cah} {\mbox{\rm [Ca/H]}}
\newcommand{\tih} {\mbox{\rm [Ti/H]}}
\newcommand{\crh} {\mbox{\rm [Cr/H]}}
\newcommand{\nih} {\mbox{\rm [Ni/H]}}
\newcommand{\znh} {\mbox{\rm [Zn/H]}}
\newcommand{\yh} {\mbox{\rm [Y/H]}}

\newcommand{\xfe} {\mbox{\rm [X/Fe]}}
\newcommand{\cfe} {\mbox{\rm [C/Fe]}}
\newcommand{\ofe} {\mbox{\rm [O/Fe]}}
\newcommand{\nafe} {\mbox{\rm [Na/Fe]}}
\newcommand{\mgfe} {\mbox{\rm [Mg/Fe]}}

\newcommand{\alfe} {\mbox{\rm [Al/Fe]}}

\newcommand{\sife} {\mbox{\rm [Si/Fe]}}
\newcommand{\sfe} {\mbox{\rm [S/Fe]}}
\newcommand{\cafe} {\mbox{\rm [Ca/Fe]}}

\newcommand{\tife} {\mbox{\rm [Ti/Fe]}}
\newcommand{\crfe} {\mbox{\rm [Cr/Fe]}}

\newcommand{\nife} {\mbox{\rm [Ni/Fe]}}

\newcommand{\znfe} {\mbox{\rm [Zn/Fe]}}
\newcommand{\yfe} {\mbox{\rm [Y/Fe]}}

\newcommand{\bafe} {\mbox{\rm [Ba/Fe]}}

\newcommand{\ymg} {\mbox{\rm [Y/Mg]}}
\newcommand{\yal} {\mbox{\rm [Y/Al]}}

\newcommand{\teff}  {\mbox{$T_{\rm eff}$}}
\newcommand{\Tc}  {\mbox{$T_{\rm c}$}}

\newcommand{\logg}  {\mbox{{\rm log}\,$g$}}

\newcommand{\turb}  {\mbox{$\xi_{\rm turb}$}}
\newcommand{\HI} {\ion{H}{i}}

\newcommand{\CI} {\ion{C}{i}}
\newcommand{\OI} {\ion{O}{i}}
\newcommand{\oI} {\mbox{\rm [{\ion{O}{i}}]}}

\newcommand{\MgI} {\ion{Mg}{i}}
\newcommand{\NaI} {\ion{Na}{i}}
\newcommand{\AlI} {\ion{Al}{i}}
\newcommand{\SiI} {\ion{Si}{i}}
\newcommand{\SI} {\ion{S}{i}}

\newcommand{\CaI} {\ion{Ca}{i}}
\newcommand{\CaII} {\ion{Ca}{ii}}
\newcommand{\TiI} {\ion{Ti}{i}}
\newcommand{\TiII} {\ion{Ti}{ii}}
\newcommand{\CrI} {\ion{Cr}{i}}
\newcommand{\CrII} {\ion{Cr}{ii}}

\newcommand{\FeI} {\ion{Fe}{i}}
\newcommand{\FeII} {\ion{Fe}{ii}}
\newcommand{\NiI} {\ion{Ni}{i}}

\newcommand{\ZnI} {\ion{Zn}{i}}

\newcommand{\YII} {\ion{Y}{ii}}
\newcommand{\BaII} {\ion{Ba}{ii}}

\def\ltsima{$\; \buildrel < \over \sim \;$}
\def\simlt{\lower.5ex\hbox{\ltsima}}
\def\gtsima{$\; \buildrel > \over \sim \;$}
\def\simgt{\lower.5ex\hbox{\gtsima}}
\begin{document}

\title{High-precision abundances of elements in {\em Kepler} LEGACY stars}

\subtitle{Verification of trends with stellar age
\thanks{Based on spectra obtained with HARPS-N@TNG under programme A33TAC\_1.}$^{,}$
\thanks{Tables 1 and 2 are also available in electronic form at the CDS via anonymous ftp
to cdsarc.u-strasbg.fr (130.79.128.5) or via http://cdsweb.u-strasbg.fr/cgi-bin/qcat?J/A+A/}}

%\author{P. E. Nissen \inst{1}}
\author{P. E. Nissen \and V. Silva Aguirre \and J. Christensen-Dalsgaard \and R. Collet
\and F. Grundahl \and D. Slumstrup} 

\institute{Stellar Astrophysics Centre, 
Department of Physics and Astronomy, Aarhus University, Ny Munkegade 120, DK--8000
Aarhus C, Denmark.  \email{pen@phys.au.dk}.}

\date{Received 28 August 2017/ Accepted 29 September 2017}

\abstract
% context heading (optional)
{A previous study of solar twin stars has revealed the existence of correlations
between some abundance ratios and stellar age providing new knowledge 
about nucleosynthesis and Galactic chemical evolution.}
% aims heading (mandatory)
{High-precision abundances of elements are determined for stars with asteroseismic ages
in order to test the solar twin relations.}
% methods heading (mandatory)
{HARPS-N spectra with signal-to-noise ratios $S/N\simgt250$ and MARCS model atmospheres
were used to derive abundances of C, O, Na, Mg, Al, Si, Ca, Ti, Cr, Fe, Ni, Zn, and Y in 
ten stars from the {\em Kepler} LEGACY sample (including the binary pair 16\,Cyg\,A and B)
selected to have metallicities in the range $-0.15<\feh<+0.15$ and
ages between 1 and 7\,Gyr. Stellar gravities were obtained from seismic data 
and effective temperatures were determined by comparing non-LTE iron abundances 
derived from \FeI\ and \FeII\ lines. Available non-LTE corrections were also applied
when deriving abundances of the other elements.}
% results heading (mandatory)
{The abundances of the {\em Kepler} stars support the \xfe -age relations previously
found for solar twins. \mgfe , \alfe , and \znfe\ decrease by $\sim \! 0.1$\,dex over the 
lifetime of the Galactic thin disk due to delayed contribution of iron
from Type Ia supernovae relative to prompt production of 
Mg, Al, and Zn in Type II supernovae. 
\ymg\ and \yal , on the other hand, increase by $\sim \! 0.3$\,dex,
which can be explained by an increasing contribution of $s$-process
elements from low-mass AGB stars as time goes on. 
The trends of \cfe\ and \ofe\ 
are more complicated due to variations  of the
ratio between refractory and volatile elements among stars of similar age. 
Two stars with about the same age as the Sun show very different trends of
\xh\ as a function of elemental condensation temperature $T_{\rm c}$ and 
for 16\,Cyg, the two components have an abundance difference,
which increases with $T_{\rm c}$. These anomalies may be connected to planet-star
interactions.}
% conclusions heading (optional)
{}


\maketitle

\section{Introduction}
\label{introduction} 
Precise determinations of abundances of elements in solar twin stars have 
revealed significant variations of some abundance ratios, for example  [Mg/Fe] and [Y/Mg],
tightly correlated with stellar age \citep{nissen15, nissen16, 
spina16a, spina16b, tuccimaia16, adibekyan16, adibekyan17}.
This is of high interest for studies of nucleosynthesis and chemical evolution
in our Galaxy. As the ages of the solar twins were derived from isochrones 
in the \teff\ - \logg\ plane and thus depend on the spectroscopically determined 
values of effective temperature and  surface gravity, there is, however, a need
to test the correlations using stellar ages determined in an independent way. 

Asteroseismology provides an alternative method to determine reliable stellar ages. 
By fitting oscillation frequencies including the large and small 
frequency separations \citep{jcd88} by stellar models,  ages can
be obtained \citep[see review by][]{jcd16}. Thus, \citet{silva-aguirre15}
derived ages of 33 exoplanet host stars from oscillation frequencies 
delivered by the {\em Kepler} mission. More recently, \citet{lund17}
have measured precise oscillation frequencies for 66 main-sequence and subgiant 
stars for which long {\em Kepler} time series are available. The data for
this so-called {\em Kepler} LEGACY sample have been analysed in
\citet{silva-aguirre17} by seven teams using different stellar
models and methods to fit the frequencies. Constraints on effective temperature,
\teff , and metallicity, \feh , were adopted from an analysis 
of high-resolution spectra with 
signal-to-noise ($S/N$) ratios of about 30 by \citet{buchhave15}. 
The resulting seismic ages were estimated to have uncertainties of typically
10\% - 20\%. Furthermore, stellar masses and radii were determined with 
uncertainties at a level of 2\% - 4\%, and surface gravities, \logg , 
with errors less than 0.01\,dex.

In this paper we have derived precise abundances for a subset of the {\em Kepler} LEGACY
stars by analysing high-resolution HARPS-N spectra with $S/N\simgt250$ and have investigated
if the relations between various abundance ratios and stellar age obtained 
for solar twins are confirmed. Based on the \citet{buchhave15} data, the stars
were selected to have \teff\ between 5700\,K and 6400\,K and \feh\ between $-0.15$
and +0.15. The metallicity range is similar to that of the solar twins
($-0.10<\feh<+0.10$), but the range in effective temperature  is much larger.
For the solar twins, an age range of about 9\,Gyr arises because of differences
in surface gravity, whereas the age range (1 - 7\,Gyr) 
of the selected LEGACY stars is primarily due to differences in \teff .

The paper is structured as follows:
In Section 2, we describe the observations and reduction of the HARPS-N spectra.
In Section 3, we are dealing with the determination of
\teff\ and abundances of elements including a discussion of non-LTE
corrections. Section 4 is devoted to a discussion of the trends
of various abundance ratios as a function of stellar age and a detailed
discussion of the binary pair, 16\,Cyg\,A and B.
Finally, in Section 5 we present a concluding summary.

\section{Observations and data reductions}
\label{observations}
The spectroscopic observations of the {\em Kepler} LEGACY stars were carried 
out July 19 - 22, 2016 with the HARPS-N spectrograph 
\citep{cosentino12} at the 3.6\,m
Telescopio Nationale Galileo (TNG) on La Palma. HARPS-N is an updated version
of the HARPS-S instrument at the ESO 3.6\,m telescope on La Silla,
which was used to obtain spectra of the solar twins analysed in 
\citet[][hereafter Paper\,I]{nissen15}.
The two spectrographs cover the wavelength range from 3800\,\AA\ to 6900\,\AA\ with a
resolution of $R = \lambda/\Delta\lambda \sim115\,000$ and a sampling of 3.3 pixels
per spectral resolution element.

Extraction of spectra,
flat fielding and wavelength calibration were done on-line with the 
HARPS-N data reduction pipeline. 
Individual spectra for a given star
were combined in IRAF after correction for Doppler shifts and normalized
to a  continuum level of $\sim \! 1.0$ as described in Paper\,I. 
Furthermore, the IRAF {\tt splot} task was used to
measure equivalent widths of the spectral lines listed in
Table 2 of Paper I by Gaussian fitting to the line profiles
relative to continuum regions lying within 3\,\AA\ from the line measured.

\begin{figure}
\resizebox{\hsize}{!}{\includegraphics{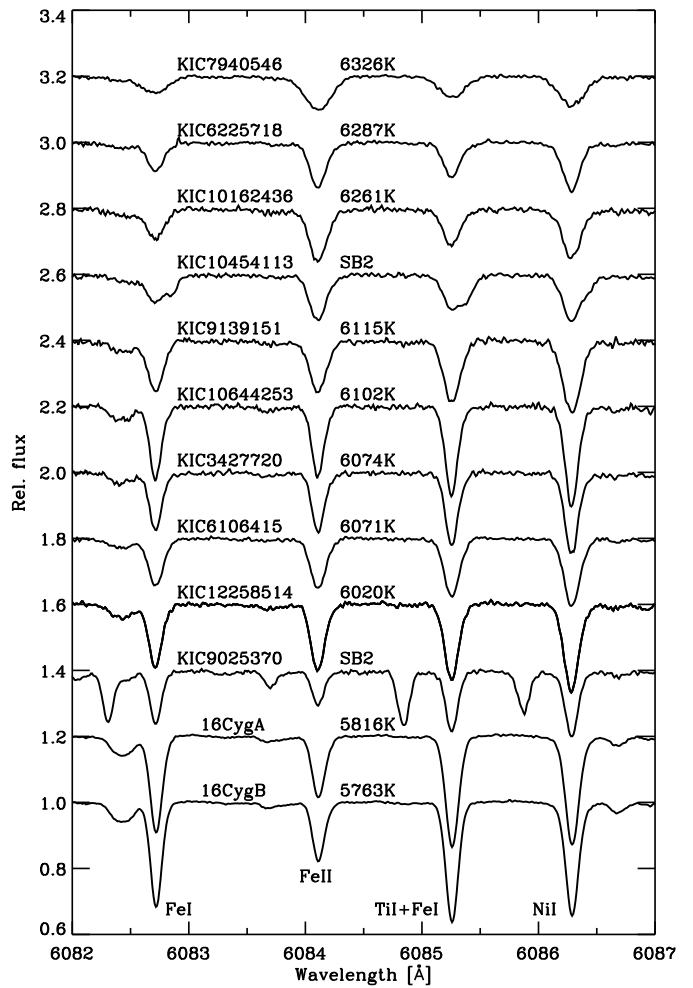}}
\caption{HARPS-N spectra in the 6082\,\AA\ -  6087\,\AA\ region 
arranged in order of increasing \teff\ and offset in successive steps of 0.2
units relative to the spectrum of 16\,Cyg\,B. The spectra are marked with the
\teff\ values determined in this paper except for the two spectroscopic
binaries.}
\label{fig:spectra}
\end{figure}

Due to poor seeing and dust in the atmosphere
only about half of the planned programme was completed, but for 12 stars
spectra with $S/N\simgt250$ were obtained.
The quality of the spectra is illustrated in Fig. \ref{fig:spectra},
which shows some typical lines used to determine abundances.
As seen the ratio of the strengths of the \FeII\ and the \FeI\ lines increases
very significantly with \teff\ over the range covered. The ratio also depends on
\logg , but since this parameter is precisely known from the seismic data,
the equivalent widths of \FeII\ lines relative to those of \FeI\
become a sensitive measure of \teff .

As seen from Fig. \ref{fig:spectra}, two stars
turn out to be double-lined spectroscopic binaries. For \object{KIC\,9025370},
two sets of well separated lines are present and for \object{KIC\,10454113}, 
the secondary component disturbs the right wings of the lines of
the primary component. These two  stars were excluded from the abundance
analysis. Secondary oscillation spectra were  not detected by \citet{lund17},
but as discussed by \citet{silva-aguirre17} asteroseismic distances of these stars
disagree with Gaia DR1 parallaxes \citep{gaia16a, gaia16b}.

As described in Sect. \ref{analysis}, stellar abundances were derived by
performing a differential analysis relative to the Sun. It is therefore 
important to have available a solar flux spectrum observed with the
same spectrograph as the stars \citep{bedell14}. We were not able
to observe such a spectrum with HARPS-N and used instead the
HARPS-S spectrum of reflected sunlight from the asteroid Vesta 
that was also used in Paper\,I to derive abundances of solar twins. 
This solar flux spectrum has a very high $S/N$ of 1200.
As the two instruments are  almost
identical, no offset between the two sets of abundances is expected, 
but as a check the bright solar twin, \object{18\,Sco} (\object{HD\,146233}),  
that were included in Paper I and have a HARPS-S spectrum with
$S/N = 850$ was observed with HARPS-N with a similar high $S/N$.
Equivalent widths measured in the two spectra for 132 spectral lines agree very well,
that is with an average deviation of $\langle \Delta$\,(N--S)$\rangle$\,=\,0.1\,m\AA\ 
and a standard deviation of $\pm 0.5$\,m\AA . Furthermore, the derived
abundances of 14 elements agree with $\langle \Delta$\,(N--S)$\rangle$\,=\,$0.001 \pm 0.006$\,dex.
This shows that the abundances in this paper are on the same system as
that of the solar twins in Paper I.

\section{Analysis}
\label{analysis}
The Uppsala EQWIDTH program together with 1D MARCS models \citep{gustafsson08} 
were used to determine elemental abundances from the measured equivalent widths
assuming local thermodynamic equilibrium (LTE). The analysis was made 
differentially line-by-line to the Sun; hence, the oscillator strengths of
the lines cancel out. Doppler broadening due to microturbulence was
specified with a parameter \turb\ added in quadrature to the thermal
broadening. Further details including references for collisional broadening 
may be found in Paper I. 

The spectral lines are also broadened by macroturbulence and stellar rotation, 
but this does not change the equivalent width of the lines, so in principle
there is no effect on the derived abundances. 
Rapid rotation makes it, however, more difficult to measure  
precise values of equivalent widths due to broader lines and more
line blending. It is not a great problem for our set of stars, for which 
the projected equatorial velocity,  $V \, {\rm sin} i$,
ranges from 1.4\,\kmprs\ to about 10\,\kmprs\ (see Table \ref{table:param}),
although some of the lines listed in Table 2 of Paper\,I
had to be excluded for the warmest and most rapidly rotating stars
due to line blending problems.

\subsection{Atmospheric parameters} 
\label{parameters}

For the surface gravity, we adopted the mean of the seven seismic \logg\ values 
derived in \citet{silva-aguirre17}.  
An effective temperature, \teff (ion), was then determined so
that the Fe abundance determined from \FeI\ lines, \fehI ,
is the same as the Fe abundance determined from \FeII\ lines, \fehII .
Furthermore, an effective temperature, \teff (exc),  
was determined by requesting that the slope of \feh\ vs. excitation 
potential for \FeI\ lines should be zero. 

\begin{table*}
\caption[ ]{List of stars with derived atmospheric parameters and seismic data.}
\label{table:param}
\centering
\setlength{\tabcolsep}{0.20cm}
\begin{tabular}{cccccccccc}
\noalign{\smallskip}
\hline\hline
\noalign{\smallskip}
  KIC no. & ID & S/N\,\tablefootmark{a} &  \teff (ion)  & \teff (exc) & \feh  &  \turb &  \logg \,\tablefootmark{b} & Age\,\tablefootmark{b} & $V {\rm sin} i$\,\tablefootmark{c}   \\
          &    &                        &   [K]         &     [K]     & [dex] &  \kmprs & [dex]                      & [Gyr] & \kmprs \\
\noalign{\smallskip}
\hline
\noalign{\smallskip}
 3427720 & \object{BD\,+38\,3428} &  280 & $6074 \pm 20$ & $6109 \pm 24$ & $-0.024 \pm 0.017$ & $1.28 \pm 0.06$ & 4.386 & $2.4 \pm 0.3$ & 4.0 \\
 6106415 & \object{HD\,177153}    &  400 & $6071 \pm 14$ & $6108 \pm 17$ & $-0.037 \pm 0.012$ & $1.29 \pm 0.04$ & 4.299 & $4.8 \pm 0.6$ & 4.0 \\
 6225718 & \object{HD\,187637}    &  370 & $6287 \pm 18$ & $6330 \pm 33$ & $-0.110 \pm 0.014$ & $1.49 \pm 0.06$ & 4.319 & $2.6 \pm 0.4$ & 5.0 \\
 7940546 & \object{HD\,175226}    &  430 & $6326 \pm 15$ & $6312 \pm 28$ & $-0.126 \pm 0.012$ & $1.71 \pm 0.05$ & 4.005 & $2.4 \pm 0.3$ & 9.7 \\
 9139151 & \object{BD\,+45\,2796} &  250 & $6115 \pm 22$ & $6158 \pm 27$ & $+0.096 \pm 0.019$ & $1.28 \pm 0.06$ & 4.379 & $1.9 \pm 0.7$ & 6.0 \\
10162436 & \object{HD\,188819}    &  250 & $6261 \pm 27$ & $6257 \pm 49$ & $-0.073 \pm 0.021$ & $1.57 \pm 0.09$ & 3.977 & $2.5 \pm 0.4$ & 6.5 \\
10644253 & \object{BD\,+48\,2683} &  250 & $6102 \pm 22$ & $6150 \pm 27$ & $+0.130 \pm 0.019$ & $1.17 \pm 0.06$ & 4.403 & $1.3 \pm 0.7$ & 3.8 \\
12069424 & \object{16\,Cyg\,A}    &  800 & $5816 \pm 10$ & $5833 \pm 12$ & $+0.093 \pm 0.007$ & $1.13 \pm 0.02$ & 4.291 & $7.0 \pm 0.5$ & 2.2 \\
12069449 & \object{16\,Cyg\,B}    &  800 & $5763 \pm 10$ & $5775 \pm 12$ & $+0.062 \pm 0.007$ & $1.06 \pm 0.02$ & 4.356 & $7.1 \pm 0.5$ & 1.4 \\
12258514 & \object{HD\,183298}    &  270 & $6020 \pm 20$ & $6071 \pm 24$ & $+0.027 \pm 0.017$ & $1.37 \pm 0.06$ & 4.124 & $4.5 \pm 0.8$ & 3.5 \\   
\noalign{\smallskip}
\hline
\end{tabular}
\tablefoot{
\tablefoottext{a}{Signal-to-noise ratio per spectral pixel at 6000\,\AA .}
\tablefoottext{b}{Seismic values of surface gravity and age from \citet{silva-aguirre17}.}
\tablefoottext{c}{Projected equatorial rotation velocity; for 16\,Cyg\,A and B seismic values from \citet{davies15}
are given, whereas spectroscopic values from \citet{bruntt12} are given for the other stars.} 
}

\end{table*}

As the derived Fe abundances depend on the microturbulence, this parameter
has to be determined together with \teff (ion) and  \teff (exc) from 
the requirement that the slope of \feh\ versus equivalent width is zero. In addition, one
must ensure that the parameters of the MARCS model applied are
consistent with those derived. Hence, the determination of \teff ,
\feh , and \turb\ is an iterative process.

In deriving the atmospheric parameters, non-LTE corrections from \citet{lind12}
were taken into account. \fehII\ is not significantly affected but
the non-LTE correction of \fehI\
increases from being negligible for 16\,Cyg\,A and B to about +0.02\,dex for stars with \teff\ around 6300\,K
and $\logg \sim 4.0$. The corresponding change of \teff (ion) is about
$-30$\,K, whereas the change of \teff (exc) is about +10\,K.

In the statistical-equilibrium calculations for iron by \citet{lind12}, 
the classical formula \citep{drawin69, steenbock84} for the
cross section of collisions 
with hydrogen atoms was adopted without any scaling factor.
Recently, \citet{amarsi16b} have made new non-LTE calculations for iron 
using quantum-mechanical cross sections for excitation and ionization  
due to collisions with hydrogen atoms. Applying these corrections for
our set of stars, we find similar small non-LTE effects as those 
based on the \citet{lind12} calculations;
the rms difference is 10\,K in both \teff (ion) and \teff (exc),
part of which may be due to numerical problems when interpolating
in the grids of corrections. Hence, the new calculations do not
change our results significantly and we prefer to keep the
\citet{lind12} corrections, which change more smoothly as a function
of \teff , \logg , and \feh\ than the \citet{amarsi16b} corrections.

The derived atmospheric parameters are given in Table \ref{table:param}.
Taking into account correlations between the parameters,
the errors have been calculated from the standard errors of 
the mean values of \fehI\ and \fehII\
(estimated from the line-to-line scatter), the errors 
of the slopes of \feh\  versus excitation potential and equivalent width, and an estimated
error of 0.01\,dex for \logg . This small error  of the
seismic gravity is justified  in  \citet{silva-aguirre17};
the uncertainties of the observed frequencies introduce an error of   
typically 0.005\,dex in \logg , and a comparison of gravities determined 
by the seven analysis teams shows a rms deviation ranging from 0.002\,dex
to 0.010\,dex for our set of stars.

\begin{figure}
\resizebox{\hsize}{!}{\includegraphics{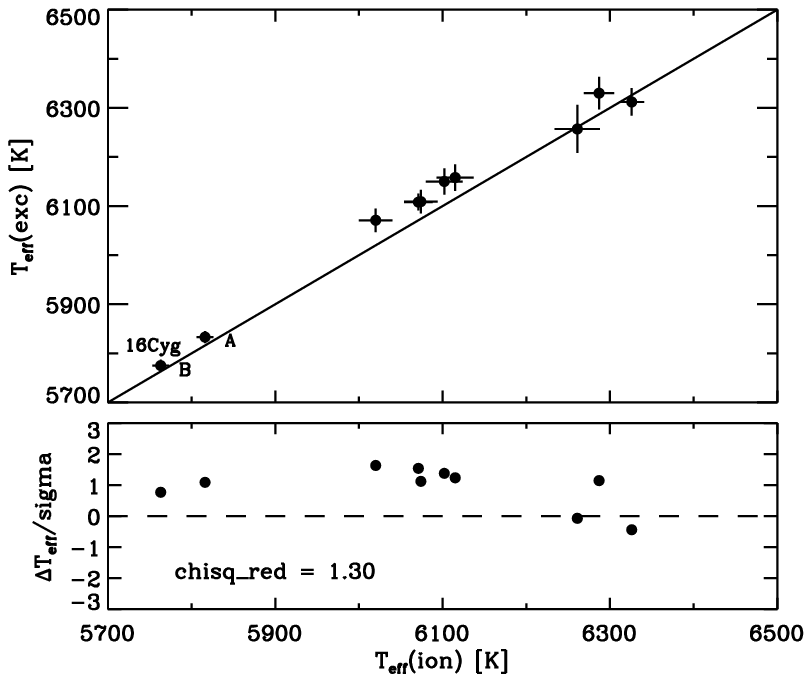}}
\caption{\teff (exc) versus \teff (ion). The lower panel shows the difference
\teff (exc) --  \teff (ion) in units of the combined error:
$\sigma = \sqrt{\sigma \teff {\rm (ion)}^2 + \sigma \teff {\rm (exc)}^2}$.}
\label{fig:Texc-Tion}
\end{figure}

In contrast to \teff (ion), \teff (exc) is not sensitive
to the adopted value of the surface gravity, because the strengths of the
\FeI\ lines used to determine the slope of \feh\ versus excitation
potential change very little with \logg . 
Therefore, a comparison of \teff (ion) and \teff (exc) provides
a test of the claimed precision of the temperatures. 
As seen from Fig. \ref{fig:Texc-Tion}, 
there is a satisfactory agreement in the sense that the deviations 
lie within about $1.8 \, \sigma$ and have a reduced chi-square of $\chi ^2 _{\rm red} = 1.3$.
There is, however, a puzzling systematic offset between
\teff (exc) and \teff (ion) for  16\,Cyg\,A and B ($\langle \Delta \rangle \, = 15$\,K) and
the five stars with \teff\ around 6100\,K ($\langle \Delta \rangle \, = 43$\,K). It is unlikely
that this is due to systematic errors in the non-LTE corrections, because for
16\,Cyg\,A and B the corrections are insignificant.

The  difference in \teff (exc) and \teff (ion) for 16\,Cyg\,A and B
disappears if the gravities are increased by approximately 0.03 dex.
It is, however, very unlikely that such a systematic error
is present in the seismic gravities.
Using the Hipparcos parallaxes for 16\,Cyg\,A and B, $47.44 \pm 0.27$\,mas and
$47.14 \pm 0.27$\,mas, respectively \citep{vanleeuwen07}, to determine
the luminosities and estimating masses by interpolation between
ASTEC evolutionary tracks \citep{jcd08a} in the $L$\,-\,\teff\ diagram,
the basic relations between gravity, mass, radius, luminosity, and effective
temperature, $g \propto M / R^2$ and $R^2 \propto L / \teff ^4$, lead to
$\logg = 4.285 \pm 0.010$ for \object{16\,Cyg\,A} and
$\logg = 4.347 \pm 0.010$ for \object{16\,Cyg\,B} in good agreement with the 
seismic gravities.

A change in the helium abundance of the  model atmospheres applied
is another possibility to
explain the discrepancy between \teff (exc) and \teff (ion).
For late-type stars, helium does not contribute to
opacity and free electrons, but it affects the mean molecular weight
and hence the pressure in the atmosphere. As discussed by \citet{bohm-vitense79}
and \citet{stromgren82}, the effect on spectral lines of a change in He abundance
is equivalent to  a change in  gravity
given by Eq. (12) in \citet{stromgren82}. The MARCS model atmospheres
have been computed with a He abundance ratio of $N_{\rm He} / N_{\rm H} = 0.085$,
which corresponds to a He mass fraction of $Y = 0.249$ close to the
solar surface value of $Y$\,=\,0.24 - 0.25 determined from
helioseismology \citep[e.g.][]{jcd91, vorontsov91, basu04}. 
If instead, the He abundance in the atmospheres of
16\,Cyg\,A and B is $Y \simeq 0.32$, the derived values of \teff (exc) and \teff (ion)
would agree. There is, however, no evidence
for enhanced He abundances in the atmospheres of 16\,Cyg\,A and B. 
From the amplitude of a small oscillatory
signature in the stellar oscillation frequencies caused by acoustic glitches
in the helium ionization zone, \citet{verma14} found the surface He abundance of 
16\,Cyg\,A and B to lie in the range $0.22 \simlt  Y \simlt 0.26$.

Hydrodynamical effects on the structure of stellar atmospheres is 
a third possibility to explain the systematic difference between 
\teff (exc) and \teff (ion).  The MARCS models assume
plane-parallel (1D) layers in hydrostatic equilibrium, whereas real
atmospheres are better represented by 3D hydrodynamical models
as shown by \citet{lind17} in a study of the 
centre-to-limb variation of Fe lines in the solar spectrum. 
A corresponding 3D non-LTE analysis of F and G stars is not available yet, but 
\citet{amarsi16b} have calculated grids of $\langle$3D$\rangle$ non-LTE corrections 
of Fe abundances derived from MARCS models, where $\langle$3D$\rangle$ refers to 
horizontally- and temporally-averaged 3D STAGGER model atmospheres 
\citep{magic13}. We have interpolated in these grids to find differential
corrections for our set of Fe lines, but this does not improve the
agreement between \teff (exc) and \teff (ion). 

It remains to be seen if a more complete 3D non-LTE analysis 
will remove the systematic offset between \teff (exc) and \teff (ion). 
For the time being, we have to accept that there could be systematic errors 
in \teff\ of the same order of size as the statistical errors quoted in 
Table \ref{table:param}.  As the errors of \teff (ion) are smaller than 
those of \teff (exc), we shall adopt \teff (ion) when deriving stellar abundances. 

\subsection{Abundances}
\label{abundances}
In this section, we describe how the abundances of elements were
derived and discuss non-LTE corrections and possible systematic errors. Statistical
errors arising from the equivalent width measurements, $\sigma _{\rm EW}$, are
estimated from the line-to-line scatter of the derived abundances. 
For 16\,Cyg\,A and B having spectra
with $S/N \sim 800$ like the solar twins, $\sigma _{\rm EW}$ was adopted
from Paper I (Table 4). For the other eight stars with
$S/N \sim 250 - 400$, we calculated
an average standard deviation $\langle \sigma \xh \rangle$ 
\citep[see Eqs. (2) and (3) in][]{nissen16}
and then the standard error of the mean value of \xh\ from
the expression $\sigma _{\rm EW}\, = \, \langle \sigma \xh \rangle / \sqrt{N_{\rm lines}}$.

\subsubsection{Non-LTE corrections and statistical errors}
\label{non-lte}

\noindent {\bf Carbon.}
The C abundances were derived from the two high-excitation 
($\chi_{\rm exc} = 7.69$\,eV) \CI\ lines at 5052.2\,\AA\ and
5380.3\,\AA . The equivalent widths of these lines range from 20\,m\AA\ to 55\,m\AA\
in our set of stars. According to \citet{takeda05a}, who adopted the \citet{drawin69}
cross sections for collisions with hydrogen atoms, differential non-LTE corrections
relative to the corrections for the Sun
are getting more negative with increasing \teff\
and decreasing gravity. The corrections 
are negligible for 16\,Cyg\,A and B,
but of some importance for the warmer stars, that is
$-0.02$\,dex for the 5052.2\,\AA\ line and $-0.01$\,dex 
for the weaker 5380.3\,\AA\ line  at $\teff \sim 6300$\,K and $\logg \sim 4.0$.
After applying the non-LTE
corrections, the C abundances derived from the two \CI\ lines
have an average standard deviation of $\langle \sigma \ch \rangle\,=\,0.017$\,dex for the 
eight stars with $\teff > 6000$\,K, corresponding to $\sigma _{\rm EW}\ch = 0.012$\,dex.

\medskip
\noindent {\bf Oxygen.}
For solar twins, O abundances were derived
from the \oI\ 6300.3\,\AA\ line taking into account a correction
of the measured equivalent width due to  a \NiI\ blend \citep{allende-prieto01}.
This \oI\ line is very weak
in our warmer stars; therefore, we have included
the \OI\ 6158.2\,\AA\ line, which is well separated from the nearby
\FeI\ 6157.7\,\AA\ line in HARPS spectra. The equivalent width of
this oxygen line increases from 3.6\,m\AA\ in the solar flux spectrum to
about 9\,m\AA\ for the warmest star, KIC\,7940546. As shown by 
\citet{bertran.de.lis15}, O abundances derived from the two oxygen lines
agree well for a large set of solar-type stars with HARPS spectra.

Recently, \citet{amarsi16a} have studied non-LTE oxygen line formation 
across a grid of 3D hydrodynamic STAGGER model atmospheres and provided
3D non-LTE corrections of O abundances determined with MARCS models. 
In contrast to the case of the \OI\ 777\,nm triplet, the corrections for
the 6158.2\,\AA\ and 6300.3\,\AA\ lines are small.
Using the IDL interpolation program provided, the differential
corrections of O abundances derived from the \OI\ 6158.2\,\AA\ line 
are found to  range from being negligible for 16\,Cyg\,A and B to
$-0.03$\,dex as the most extreme value for our sample of stars. 
In the case of the \oI\ 6300.3\,\AA\ line,
the most extreme differential correction is only $-0.01$\,dex.
This small correction arises entirely from 3D effects, because
LTE is a very good approximation
for this forbidden line \citep[e.g.][]{kiselman93}.

The oxygen abundances derived from the two \OI\ lines agree
with an average standard deviation of $\langle \sigma \oh \rangle = 0.032$\,dex.
This is higher than in the case of carbon, which is due to
larger relative errors for the equivalent widths of the weak oxygen lines. 
For one star, \object{KIC\,10162436}, only the \OI\ 6158.2\,\AA\ line
could be used, because the \oI\ 6300.3\,\AA\ line is blended by a telluric 
O$_2$ line.

\medskip
\noindent {\bf Sodium.}
The Na abundances were derived from the \NaI\ 6154.2, 6160.8\,\AA\
doublet. The equivalent widths of these lines range from about 20\,m\AA\ to 60\,m\AA\ 
in our spectra.  Non-LTE corrections were adopted from \citet{lind11}
using the available IDL program to interpolate to the atmospheric parameters
of the stars. The corrections of \nah\ are negative reaching
about $-0.015$\,dex for the warmest stars in our sample. Although small, these
corrections are not negligible in comparison with the statistical error of \nah ;
from the comparison of the Na abundances derived from the two lines we find
an average standard deviation of $\langle \sigma \nah \rangle = 0.012$\,dex.

\medskip
\noindent {\bf Magnesium.}
The Mg abundances were obtained from the 4730.0\,\AA\ and
5711.1\,\AA\  \MgI\ lines of which the first have
equivalent widths from 40\,m\AA\
to 80\,m\AA\ in our spectra, whereas the second line ranges
from 80\,m\AA\ to 120\,m\AA . In order to obtain precise equivalent widths,
this stronger line was measured with Voigt profile fitting
instead of Gauss fitting.

Non-LTE corrections for the 5711.1\,\AA\  \MgI\ line were adopted  from
\citet{osorio16}, who applied new 
quantum-mechanical calculations of cross sections for collisions with
hydrogen \citep{barklem12} and electrons \citep{osorio15}.
The interactive database
INSPECT\,\footnote{{\tt http://www.inspect-stars.com}} was used to interpolate 
to the stellar parameters resulting in corrections of \mgh\
varying from being negligible for 16\,Cyg\,A and B to about +0.02\,dex
for the warmest stars. Non-LTE corrections for the 4730.0\,\AA\ \MgI\
line are not available at INSPECT, but may be obtained with
{\tt Spectrum Tools}\,\footnote{{\tt http://nlte.mpia.de}}. The non-LTE data for Mg
at this site are based on the model atoms and statistical-equilibrium
calculations described in \citet{bergemann15, bergemann17}, who also adopted
quantum-mechanical hydrogen collisional rates from \citet{barklem12},
but rates of electron collisions were calculated with classical formulae.
Still, the differential non-LTE
corrections of abundances derived from the 5711.1\,\AA\  \MgI\ line
agree almost exactly with those of \citet{osorio16}. The corrections
of abundances obtained from the weaker 4730.0\,\AA\ line are about half
the corrections for the 5711.1\,\AA\ line. 

The abundances derived from the two \MgI\ lines show good
agreement, that is $\langle \sigma \mgh \rangle = 0.015$\,dex. 

\medskip
\noindent {\bf Aluminium.}
The \AlI\ 6696.0, 6698.7\,\AA\ doublet was used to derive Al abundances.
In our spectra, the equivalent widths of these lines range from about 10 to 45\,m\AA\
with the 6698.7\,\AA\ line being weaker than the 6696.0\,\AA\ line
by about a factor of two. Non-LTE corrections were adopted from the 
statistical-equilibrium study by \citet{nordlander17}, who applied 
inelastic hydrogen-collisional rates of \citet{belyaev13}
for the six lower-excitation states of \AlI . The 
corrections of \alh\ are negligible for 16\,Cyg\,A and B but rise to
about +0.040\,dex (for both lines) for our warmest stars. 

Non-LTE corrections have also  been published by \citet[][Table 2]{mashonkina16}.
Their data show a steeper
rise of the corrections as a function of increasing \teff\ leading to a
non-LTE correction of +0.08\,dex for our warmest star. It is unclear what is causing
the difference relative to Nordlander \& Lind. 
Mashonkina et al. also applied the \citet{belyaev13} cross sections,
but since Nordlander \& Lind used more realistic calculations for electron collisions
than the formula of \citet{regemorter62} adopted by Mashonkina et al., we shall
use the Nordlander-Lind corrections. 

The comparison of Al abundances derived from the two \AlI\ lines
results in an average standard deviation of 0.016\,dex for the  eight stars
with $\teff > 6000$\,K, nearly the same value
as obtained for the C and Mg lines, but higher than in the case of Na.

\medskip
\noindent {\bf Silicon.}
The Si abundances were determined from 7 - 10
high-excitation ($\chi_{\rm exc}$ = 4.0 - 6.0\,eV) \SiI\ lines
with equivalent widths between 8 and 55\,m\AA\ in our spectra.
By comparing Si abundances derived from different lines we 
obtain $\langle \sigma \sih \rangle = 0.020$\,dex.

For the \SiI\ lines at 5645.6\,\AA\ and 5665.6\,\AA ,
{\tt Spectrum Tools} provides non-LTE corrections based on the 
statistical-equilibrium calculations of \citet{bergemann13}
\citep[see also][]{bergemann14}. The
differential corrections relative to the Sun are very small for 
our set of stars, at most $-0.002$\,dex. As the other \SiI\ lines
have similar excitation potentials and equivalent widths, the non-LTE corrections are
probably also small for these lines.

The calculations of  \citet{bergemann13} are based on the semi-classical
formula of \citet[][Eq. A10]{lambert93} for the hydrogen-collisional rate.
\citet{mashonkina16} have applied more advanced data from
\citet{belyaev14} to calculate non-LTE corrections for \SiI\ lines. They do not
give numbers but state that the corrections are ``vanishingly small`` 
for solar-type stars.
Furthermore, \citet{amarsi17} have applied the quantum-mechanical
collisional rates of \citet{belyaev14} in a 3D non-LTE study of the Si
abundance in the solar atmosphere. Several of our lines are included, and
in agreement with \citet{bergemann13} they find a non-LTE correction
of $-0.01$\,dex only. Altogether, it seems safe to assume that we can neglect
the differential non-LTE \SiI\ corrections for our sample of stars.

\medskip
\noindent {\bf Sulphur.}
For the solar twins, S abundances were determined from four high-excitation
($\chi_{\rm exc} = 7.87$\,eV) \SI\ lines,
which range in equivalent width from about 10 to 30\,m\AA\ in our spectra.
For the warmer stars, the abundance derived from the 6046.0\,\AA\ line has
a systematic deviation of about $-0.10$\,dex from the mean abundance of the other three lines.
This is probably due to an unidentified blend in the line, which was  therefore
omitted. For the remaining three lines, the average standard deviation 
of the derived abundances
is $\langle \sigma \sh \rangle = 0.033$\,dex. This relatively high value
can be ascribed to the weakness of the sulphur lines and difficulties in 
setting the continuum level. Still, the error of
the mean value of [S/H] has a satisfactory low value of 
$\sigma _{\rm EW}\sh = 0.019$\,dex.

According to \citet{takeda05b}, the non-LTE corrections for the 6052.7\,\AA\
and 6757.1\,\AA\ \SI\ lines are small (i.e. about $-0.01$ to $-0.02$\,dex) and do not
depend significantly on the atmospheric parameters of our stars. Therefore, 
differential corrections of S abundances relative to the Sun can be neglected.
The 6743.6\,\AA\ line is not included by  \citet{takeda05b}, but since this line belong
to the same multiplet as the 6757.1\,\AA\ line, the
non-LTE corrections are also expected so be negligible. We note, however, that
the statistical-equilibrium calculations of \citet{takeda05b} are based
on cross sections for hydrogen collisions according to the Drawin formula.
An update of the computations to quantum-mechanical cross sections would
be desirable.
 
\medskip
\noindent {\bf Calcium.}
Non-LTE calculations for Ca lines in spectra of late-type stars 
based on quantum-mechanical rates  for \CaI\ + \HI\ collisions have been
carried out by \citet{mashonkina17}. Four of our seven \CaI\ lines (5582.0\,\AA ,
5590.1\,\AA ,  6166.4\,\AA , and 6455.6\,\AA ) are included. The average of
the differential non-LTE corrections relative to the Sun are small, 
ranging from $-0.007$\,dex to +0.001\,dex for our set of stars.
Assuming that the non-LTE corrections for the other three \CaI\ lines (5260.4\,\AA , 5513.0\,\AA , and 
5867.6\,\AA ) are also  negligible small, it turns out that the 5513.0\,\AA\ line
leads to systematically deviating abundances
for stars with $\teff > 6000$\,K probably because a line from another
element coincides with the \CaI\ line. Omitting this line, the Ca abundances derived
from the other lines agree with an average standard deviation of 
$\langle \sigma \cah \rangle = 0.026$\,dex.

\medskip
\noindent {\bf Titanium.}
Ti abundances were derived from 11 \TiI\ and three \TiII\ lines.
For the solar twins the  equivalent widths range from about 10 to 60\,m\AA ,
which is ideal for precise
abundance determinations, but as the \TiI\ lines have low excitation potentials,
several lines are very  weak in the spectra of the warmest stars 
making the derived abundances uncertain.
After excluding these lines, the average standard deviation becomes 0.028\,dex
for the \TiI\ lines, whereas it is 0.019\,dex for the three \TiII\ lines

For all \TiI\ lines, {\tt Spectrum Tools} provides non-LTE corrections based
on the statistical-equilibrium calculations of \citet{bergemann11}. The corrections
of $\tih_{\rm \TiI}$ increase with increasing \teff\ and decreasing \logg\ 
reaching about +0.03\,dex
for our warmest and most evolved stars (\object{KIC\,7940546} and \object{KIC\,10162436}).
According to \citet{bergemann11}, non-LTE corrections for the \TiII\ lines
are negligible and it is therefore interesting to compare Ti abundances
derived from \TiI\ and \TiII\ lines. Assuming LTE, we get a mean difference
$\langle \tih_{\rm \TiI} - \tih_{\rm \TiII} \rangle\, = -0.013 \pm 0.006$. 
Including the non-LTE corrections, the difference becomes
$\langle \tih_{\rm \TiI} - \tih_{\rm \TiII} \rangle \, = +0.002 \pm 0.004$, that is a
better agreement. Considering the errors, this is not
a proof of the validity of the non-LTE corrections, but we conclude that
there is a very satisfactory agreement between abundances derived from
\TiI\ and \TiII\ lines supporting the \teff\ values derived by      
requesting that Fe abundances derived from \FeI\ and \FeII\ lines
should  be the same when adopting the seismic  \logg\ value
(see Sect. 3.1).

\medskip
\noindent {\bf Chromium.}
Cr abundances were derived from 6 - 8 \CrI\ and two \CrII\ lines.
There is no systematic calculations of non-LTE corrections as a function of
\teff\ and \logg\ for solar-type stars, but
\citet{bergemann10}  have presented a non-LTE study of Cr abundances in the Sun
and some metal-poor stars. Neglecting collisions with hydrogen atoms,
they found quite large non-LTE corrections for the Sun, that is +0.05 to +0.10\,dex
for our \CrI\ lines, whereas the corrections for the \CrII\ lines are negligible.

Although new  statistical-equilibrium calculations including 
hydrogen collisions according to the Drawin formula  
by \citet{scott15a} lead to smaller corrections for
solar \CrI\ lines ($\sim \! 0.025$\,dex), differential corrections relative to
the Sun may 
be significant for our warmer stars. To check this we have compared Cr abundances
derived from \CrI\ and \CrII\ lines. For 16\,Cyg\,A and B there is almost perfect 
agreement, but for the other stars we find a mean difference  
$\langle \crh_{\rm \CrI} - \crh_{\rm \CrII} \rangle \, = -0.014 \pm 0.007$. Assuming this
is due to non-LTE effects, we have added 0.014\,dex
to the abundances derived from \CrI\ lines and calculated a mean Cr abundance
from the \CrI\ and \CrII\ lines giving equal weight to all lines. 
The resulting average standard deviation of \crh\  is 0.029\,dex.

\medskip
\noindent {\bf Iron.}
As discussed in Sect. \ref{parameters}, the non-LTE corrections of \citet{lind12}  
are negligible for \FeII\ lines but reach about +0.02\,dex for \FeI\ lines
in the spectra of our warmest stars. Because effective temperatures
were derived so that the same Fe abundance is obtained from \FeI\ and \FeII\ lines,
and because there are many more \FeI\ lines ($N_{\rm lines}$ = 40 to 47) than \FeII\ lines
($N_{\rm lines}$ = 8 or 9), the error of \feh\  arising from the equivalent width 
measurements gets the largest
contribution from the \FeII\ lines. The average standard deviation 
is nearly the same for \FeI\ and \FeII\ lines, 0.021\,dex and 0.022\,dex, respectively. 

\medskip
\noindent {\bf Nickel.}
The Ni abundances were determined from 10 to 14 \NiI\ lines
having equivalent widths between 5 and 100\,m\AA . From the the line-to-line scatter of \nih\
we find an average standard deviation of 0.021\,dex, which is the same as obtained
for the \FeI\ lines.

Unfortunately, there are no  non-LTE calculations for nickel. In their
determination of the photospheric Ni abundance for the Sun, 
\citet{scott15a} argue that non-LTE  
corrections are small, which is supported by the fact that they find agreement
between the derived solar abundance, $A({\rm Ni}) = 6.20 \pm 0.04$, and the meteoritic
abundance $A({\rm Ni}) = 6.20 \pm 0.01$ \citep{lodders09}. However, it cannot be excluded
that there are small non-LTE corrections on the Ni abundances of the
{\em Kepler} stars. \NiI\ has about the same ionization potential
and a similar energy level structure as 
\FeI , so it would not be surprising if the non-LTE corrections are of the same order of size.

\medskip
\noindent {\bf Zinc.} 
Zn abundances were derived from two medium-strong \ZnI\ lines at
4722.2\,\AA\ and 4810.5\,\AA\ and a weak \ZnI\ line at 6362.4\,\AA .
Non-LTE corrections with hydrogen-collisional rates according to the Drawin
formula have been calculated by \citet{takeda05b}. 
The correction of \znh\ reaches a maximum of about +0.03\,dex for the stronger
lines, but only +0.01\,dex for the 6362.4\,\AA\ line. After applying these corrections,
the average standard deviation of the abundances derived from the three lines
is 0.033\,dex. This is higher than for most elements,
which is probably due to difficulties in setting the continuum level for the blue lines.

\medskip
\noindent {\bf Yttrium.} 
Three \YII\ lines at 4883.7, 5087.4, and 5200.4\,\AA\ 
were used to determine Y abundances.  The equivalent widths of the strongest line (4883.7\,\AA )
range from 60\,m\AA\ to 74\,m\AA\ and those of the weakest line 
(5200.4\,\AA ) from 38\,m\AA\ to 46\,m\AA\ for our sample of stars.
Non-LTE studies are not available, but since the single-ionized state 
is by far the most populated,
we expect corrections for \YII\ lines to be small
as in the case of the \TiII , \CrII , and \FeII\ lines, but this should be
investigated in more detail.

Yttrium is an odd-$Z$ element, so hyperfine structure (HFS) is present
in the spectral lines. However, as discussed by \citet{hannaford82}
in their determination of the solar Y abundance,
the HFS splitting is less than 1\,m\AA . This is a factor 3 to 4 smaller
than the Doppler width of the \YII\ lines due to thermal broadening 
and microturbulence. Therefore, we have neglected HFS effects when deriving \yh . 

The Y abundances derived from the three lines agree very well; the average
standard deviation of \yh\ is 0.016\,dex close to that of Mg and Al.
This high precision is important, because  the Y/Mg and Y/Al ratios are
applied to study the Galactic evolution of a typical $s$-process element
as discussed in Sect. \ref{discussion}.

\begin{figure*}
\resizebox{\hsize}{!}{\includegraphics{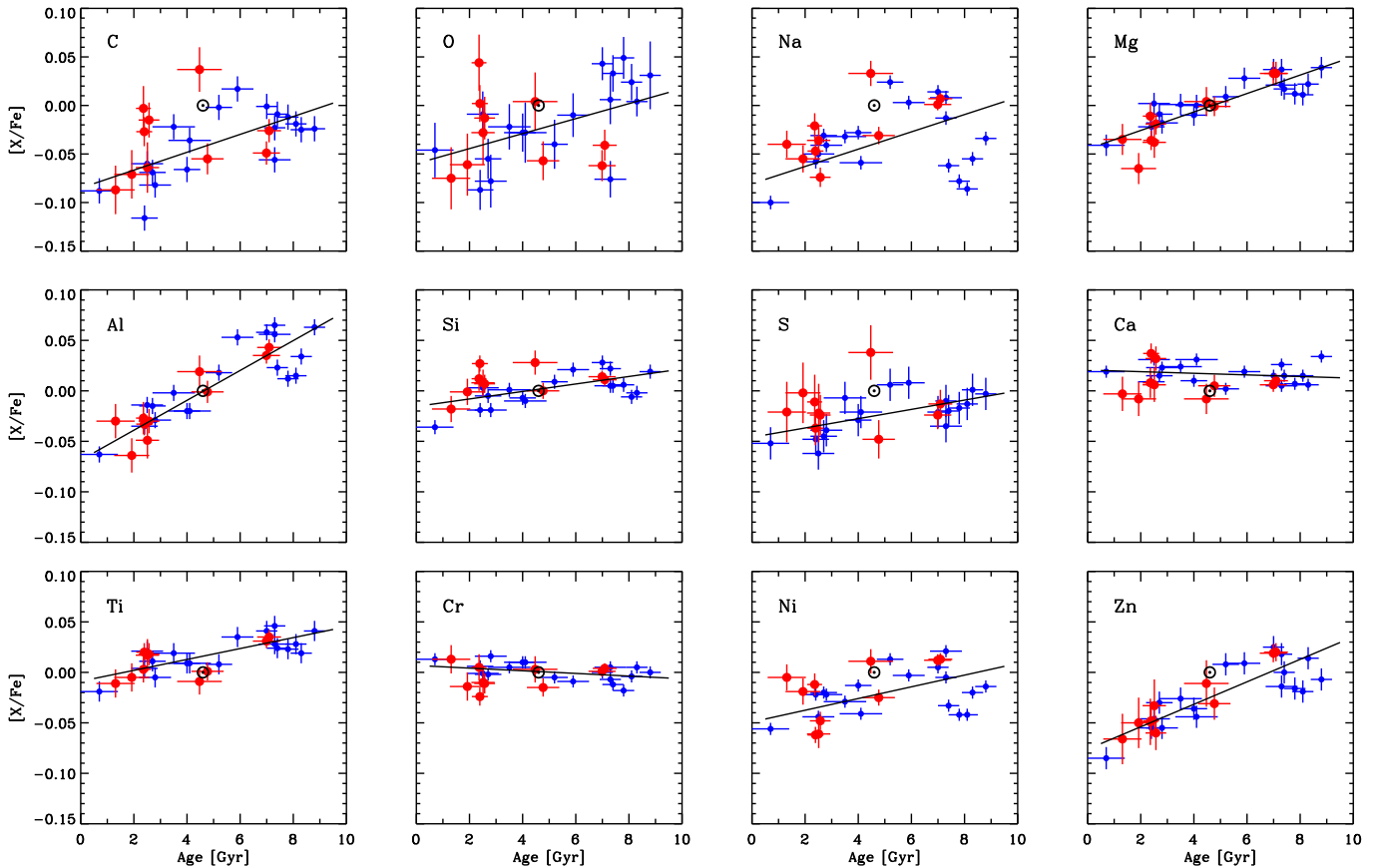}}
\caption{Abundance ratios \xfe\ as a function of stellar age.  Large (red)
filled circles refer to {\em Kepler} LEGACY stars and small (blue) filled
circles to solar twins from \citet{nissen15}.
The lines show maximum likelihood linear fits to the data with
errors in both coordinates taken into account. The Sun, shown with the
$\odot$ symbol, was not included in the fits.
Zero points and slope coefficients  for the fits are given in Table \ref{table:fits}.}
\label{fig:xfe-age}
\end{figure*}

\subsubsection{3D effects and systematic errors}
\label{3D}
Studies of non-LTE effects on derived stellar abundances
have progressed considerably in recent years as discussed above.
Still, it is only in the case of oxygen that full 3D non-LTE calculations
on an extensive grid of stellar parameters
have been carried out \citep{amarsi16a}. Although the 3D non-LTE corrections
for the \oh\ abundances derived from the 6158.2\,\AA\ and 6300.3\,\AA\ lines are
small for our sample of stars, it could be that full 3D non-LTE calculations
for the other elements will lead to significant changes of the derived
abundances.

An associated problem is the microturbulent broadening of spectral lines. 
For the 1D models, we assume that it can be described by a single parameter
determined so that
\feh\ from \FeI\ lines has no systematic dependence on equivalent width,
but in real stellar atmospheres, the microturbulence probably depends on depth. 
If element abundances are based on weak lines
on the linear part of the curve of growth (C, O, Na, Al, Si, and S in our case)
this is not a significant problem, but if stronger lines on the flat part of the
curve of growth are involved (Mg, Ca, Ti, Cr, Ni, Zn, and Y) a variation
of microturbulence with atmospheric depth  may affect the derived \xfe\ values.
Hydrodynamical models will take care of this problem, but 3D non-LTE studies
are computationally very demanding and beyond the scope
of this paper. As further discussed in Sect. \ref{xfe-age} we may, however, get an
empirical check of the importance of the 3D effects by comparing
\xfe\ values for stars having similar ages but different effective temperatures; 
as \xfe\ is measured relative to the Sun, one would  expect to see the largest
3D effects for the warmest stars in our sample.

Variations in stellar activity among stars of similar age is another
potential problem when determining high-precision stellar abundances.
\citet{flores16} have discovered that one 
of the solar twins included in Paper\,I, \object{HD\,45184}, has a 5.14\,yr activity
cycle over which the equivalent widths of strong \FeII\ lines change with
an amplitude of about $\pm 1$\,m\AA , and \citet{reddy17} have shown
that derived \bafe\ values of young solar twins are positively correlated with the \CaII\ stellar
activity index. They suggest that the very high \bafe\ values derived for stars
in young open clusters \citep{dorazi09} are not real but due to an underestimation
of the microturbulence in the upper photospheric layers of active stars,
where the \BaII\ 5853.7\,\AA\ line used to determine Ba abundances
is formed. Although our \YII\ lines are
weaker than the \BaII\ 5853.7\,\AA\ line and formed in deeper layers,
we cannot exclude that this problem also has some effect on the derived  yttrium abundances
and hence on the \ymg \,-\,age relation discussed in Sect. \ref{xfe-age}.

\begin{table*}
\caption[ ]{\feh\ and \xfe\ for 13 elements.}
\label{table:abun}
\setlength{\tabcolsep}{0.11cm}
\begin{tabular}{lrcrcrcrcrcrcrc}
\noalign{\smallskip}
\hline\hline
\noalign{\smallskip}
 ID & \feh &$\sigma$ & \cfe &$\sigma$ & \ofe &$\sigma$ & \nafe &$\sigma$& \mgfe &$\sigma$ & \alfe &$\sigma$ & \sife &$\sigma$  \\
\noalign{\smallskip}
\hline
\noalign{\smallskip}
\object{KIC\,3427720} &  $-$0.024 & 0.017 &  $-$0.003 & 0.023 &   0.044 & 0.029 &  $-$0.021 & 0.013 &  $-$0.011 & 0.014 &  $-$0.027 & 0.016 &   0.012 & 0.012\\
\object{KIC\,6106415} &  $-$0.037 & 0.012 &  $-$0.055 & 0.016 &  $-$0.057 & 0.020 &  $-$0.031 & 0.009 &  $-$0.001 & 0.010 &  $-$0.001 & 0.011 &   0.000 & 0.008\\
\object{KIC\,6225718} &  $-$0.110 & 0.014 &  $-$0.015 & 0.018 &  $-$0.013 & 0.022 &  $-$0.074 & 0.010 &  $-$0.019 & 0.011 &  $-$0.030 & 0.012 &   0.007 & 0.009\\
\object{KIC\,7940546} &  $-$0.126 & 0.012 &  $-$0.027 & 0.015 &   0.002 & 0.019 &  $-$0.047 & 0.009 &  $-$0.036 & 0.010 &  $-$0.033 & 0.011 &   0.027 & 0.008\\
\object{KIC\,9139151} &   0.096 & 0.019 &  $-$0.071 & 0.025 &  $-$0.061 & 0.032 &  $-$0.055 & 0.014 &  $-$0.065 & 0.016 &  $-$0.064 & 0.017 &  $-$0.001 & 0.013\\
\object{KIC\,10162436} &  $-$0.073 & 0.021 &  $-$0.064 & 0.026 &  $-$0.028 & 0.033 &  $-$0.036 & 0.015 &  $-$0.038 & 0.016 &  $-$0.049 & 0.018 &   0.008 & 0.013\\
\object{KIC\,10644253} &   0.130 & 0.019 &  $-$0.087 & 0.025 &  $-$0.075 & 0.032 &  $-$0.040 & 0.014 &  $-$0.035 & 0.016 &  $-$0.030 & 0.017 &  $-$0.018 & 0.013\\
\object{16\,Cyg\,A}    &   0.093 & 0.007 &  $-$0.049 & 0.012 &  $-$0.062 & 0.016 &   0.001 & 0.006 &   0.033 & 0.012 &   0.035 & 0.008 &   0.014 & 0.006\\
\object{16\,Cyg\,B}    &   0.062 & 0.007 &  $-$0.026 & 0.012 &  $-$0.041 & 0.016 &   0.007 & 0.006 &   0.033 & 0.012 &   0.043 & 0.008 &   0.011 & 0.006\\
\object{KIC\,12258514} &   0.027 & 0.017 &   0.037 & 0.023 &   0.004 & 0.030 &   0.033 & 0.013 &   0.004 & 0.015 &   0.019 & 0.016 &   0.028 & 0.012\\
\noalign{\smallskip}
\noalign{\smallskip}
\hline\hline
\noalign{\smallskip}
  ID & \sfe &$\sigma$ & \cafe &$\sigma$ & \tife &$\sigma$ & \crfe &$\sigma$& \nife &$\sigma$ & \znfe &$\sigma$ & \yfe &$\sigma$ \\
\noalign{\smallskip}
\hline
\noalign{\smallskip}
KIC\,3427720 &  $-$0.011 & 0.027 &   0.008 & 0.015 &   0.003 & 0.013 &   0.005 & 0.013 &  $-$0.012 & 0.011 &  $-$0.049 & 0.023 &  $-$0.006 & 0.015\\
KIC\,6106415 &  $-$0.048 & 0.019 &   0.005 & 0.011 &   0.001 & 0.009 &  $-$0.015 & 0.009 &  $-$0.025 & 0.008 &  $-$0.031 & 0.016 &   0.001 & 0.011\\
KIC\,6225718 &  $-$0.024 & 0.020 &   0.032 & 0.012 &   0.018 & 0.011 &  $-$0.011 & 0.010 &  $-$0.048 & 0.009 &  $-$0.060 & 0.017 &   0.032 & 0.012\\
KIC\,7940546 &  $-$0.037 & 0.017 &   0.037 & 0.010 &   0.020 & 0.009 &  $-$0.024 & 0.009 &  $-$0.062 & 0.008 &  $-$0.048 & 0.015 &   0.020 & 0.010\\
KIC\,9139151 &  $-$0.002 & 0.030 &  $-$0.008 & 0.017 &  $-$0.005 & 0.014 &  $-$0.014 & 0.014 &  $-$0.019 & 0.013 &  $-$0.050 & 0.025 &   0.036 & 0.017\\
KIC\,10162436 &  $-$0.022 & 0.029 &   0.006 & 0.017 &   0.017 & 0.016 &  $-$0.010 & 0.015 &  $-$0.061 & 0.014 &  $-$0.033 & 0.026 &   0.033 & 0.018\\
KIC\,10644253 &  $-$0.021 & 0.030 &  $-$0.003 & 0.017 &  $-$0.011 & 0.014 &   0.013 & 0.014 &  $-$0.005 & 0.013 &  $-$0.066 & 0.025 &   0.031 & 0.017\\
16\,Cyg\,A &  $-$0.024 & 0.014 &   0.006 & 0.007 &   0.031 & 0.007 &   0.001 & 0.005 &   0.012 & 0.006 &   0.019 & 0.009 &  $-$0.057 & 0.007\\
16\,Cyg\,B &  $-$0.013 & 0.014 &   0.010 & 0.007 &   0.035 & 0.007 &   0.004 & 0.005 &   0.013 & 0.006 &   0.020 & 0.009 &  $-$0.052 & 0.007\\
KIC\,12258514 &   0.038 & 0.027 &  $-$0.008 & 0.015 &  $-$0.009 & 0.013 &   0.003 & 0.013 &   0.011 & 0.012 &  $-$0.011 & 0.023 &  $-$0.047 & 0.016\\
\noalign{\smallskip}
\hline

\end{tabular}

\end{table*}

\section{Discussion}
\label{discussion}
The derived \xfe\ values of the {\em Kepler} stars are given in Table \ref{table:abun}.
The listed errors are a combination of the errors arising from the
measurement of equivalent widths as estimated from the line-to-line scatter,
(see Sect. \ref{non-lte}) and the errors arising from the uncertainties in 
\teff , \logg , \feh , and \turb .

Stellar ages from \citet{silva-aguirre17} are given in Table \ref{table:param}.
For each star, the age is calculated as the mean of the results from the seven different
analyses of the available seismic data. The age error
is calculated as a quadratic combination of the mean of the errors
estimated for the seven analyses and the standard deviation 
resulting from a comparison of ages of the different analyses.
In this way, we take into account both the statistical error due to
uncertainties of the measured oscillation frequencies and the systematic
uncertainty arising from the analysis method.

The ages of the {\em Kepler} LEGACY stars derived in \citet{silva-aguirre17}
are primarily based on the observed oscillation frequencies but constraints on
\teff\ and \feh\ are also included in the analysis. 
These parameters were taken from \citet{buchhave15}
and assumed to have errors of $\pm 77$\,K and $\pm 0.1$\,dex, respectively, except
in the case of 16\,Cyg\,A and B for which more precise values from  
\citet{ramirez09} were adopted.
For all stars our metallicities agree with those used in \citet{silva-aguirre17} within
their one-sigma uncertainty, but for three stars, 
\object{KIC\,7940546}, \object{KIC\,9139151}, and \object{KIC\,10162436}, 
our \teff\ values deviate with 91\,K, $-187$\,K, and 115\,K, respectively.
In order to see if the new values of \teff\ and \feh\
have any significant impact on the ages derived, we have repeated
the age determinations for the ASTFIT \citep{jcd08b, jcd08a} and
BASTA  \citep{silva-aguirre15} analysis methods. In both cases, the improved spectroscopic 
parameters lead to age changes having an rms deviation of only 0.2\,Gyr relative to
the ages determined in  \citet{silva-aguirre17}, and for all stars
the age change is less than the age error given in Table \ref{table:param}.

\subsection{\xfe - age correlations}
\label{xfe-age}
Fig. \ref{fig:xfe-age} shows \xfe\ as a function of stellar age
for the ten {\em Kepler} LEGACY stars  and 18 solar twins
from Paper I (excluding three $\alpha$-enhanced stars).
For the solar twins, we have adopted 
isochrone ages determined from evolutionary tracks in the \logg -\teff\
diagram calculated with the Aarhus Stellar Evolution Code, ASTEC
\citep[see][]{nissen16}.
Linear fits ($\xfe = a + b \,\cdot\,Age$) to all data 
were determined by a maximum likelihood program that includes errors in both
coordinates \citep{press92}. The coefficients $a$ and $b$
and the reduced chi-squares of the fits are given in Table \ref{table:fits}.

\begin{table}
\caption[ ]{Linear fits of \xfe\ as a function of stellar age.}
\label{table:fits}
\centering
\setlength{\tabcolsep}{0.20cm}
\begin{tabular}{rccr}
\noalign{\smallskip}
\hline\hline
\noalign{\smallskip}
       & $a$   & $b$                        & $\chi^2_{\rm red}$ \\
       & [dex] & $10^{-3}$\,dex\,Gyr$^{-1}$ &                    \\
\noalign{\smallskip}
\hline
\noalign{\smallskip}
 \cfe  &  $-0.085\,\pm 0.009$ & $ +9.2\,\pm 1.7$ &  3.6 \\
 \ofe  &  $-0.060\,\pm 0.013$ & $ +7.7\,\pm 2.4$ &  3.1 \\
 \nafe &  $-0.081\,\pm 0.016$ & $ +8.9\,\pm 2.5$ & 16.8 \\
 \mgfe &  $-0.045\,\pm 0.004$ & $ +9.5\,\pm 0.7$ &  1.1  \\
 \alfe &  $-0.069\,\pm 0.007$ & $+14.8\,\pm 1.1$ &  2.7  \\
 \sife &  $-0.015\,\pm 0.004$ & $ +3.7\,\pm 0.7$ &  3.0  \\
 \sfe  &  $-0.046\,\pm 0.006$ & $ +4.6\,\pm 1.0$ &  1.0  \\
 \cafe &  $+0.021\,\pm 0.003$ & $ -0.8\,\pm 0.3$ &  2.1  \\
 \tife &  $-0.009\,\pm 0.004$ & $ +5.4\,\pm 0.6$ &  1.1  \\
 \crfe &  $+0.007\,\pm 0.003$ & $ -1.3\,\pm 0.5$ &  1.8  \\
 \nife &  $-0.049\,\pm 0.008$ & $ +5.8\,\pm 1.2$ &  8.9  \\
 \znfe &  $-0.077\,\pm 0.006$ & $+11.2\,\pm 0.9$ &  1.8  \\
\noalign{\smallskip}
\hline
\end{tabular}
\end{table}

As seen from Fig. \ref{fig:xfe-age}, the {\em Kepler} stars
have similar dependencies of \xfe\ on age  as the solar twins, 
albeit with a larger scatter due to larger errors of \xfe .
We also note that the deviations of \xfe\ from the fitted lines
do not have any significant dependence on \teff , 
suggesting that 3D effects are not important for the
derived age trends (see Sect. \ref{3D}). 

Among the element ratios shown in Fig. \ref{fig:xfe-age},
\mgfe ,  \alfe  , and \znfe\ have particular tight trends
as a function of stellar age. 
Over the lifetime of the Galactic disk, these abundance
ratios decrease by about 0.1\,dex. Given that Mg, Al, and Zn are
exclusively made in Type II supernovae (SNe), whereas Fe is also
produced by Type Ia SNe \citep[e.g.][]{kobayashi06}, this can be explained as
due to an increasing number of Type Ia SNe relative to the number
of Type II SNe in the course of time. 
\sife , \sfe , and \tife\ also increase with stellar age, but
the amplitude of the variations is smaller ($\sim 0.05$\,dex), because
Si, S, and Ti have some contribution from Type Ia SNe
in addition to the Type II SNe contribution \citep{kobayashi06}. 
\cafe\ and \crfe , on the other hand, are nearly constant with age. 
While this is expected
for \crfe , the constancy of \cafe\ is surprising.
As discussed in Paper I, the explanation may be that low luminosity
SNe producing large amounts of Ca \citep{perets10} are important
for the chemical evolution of Ca.

\cfe , \ofe , \nafe , and \nife\ also increase as a function of age,
but the scatter is larger than expected from the estimated errors
of \xfe\ and age, that is $\chi^2_{\rm red} > 3$ as seen from Table \ref{table:fits}.
In the case of Na and Ni, the large scatter is mainly due to the existence of five 
old stars with low  values of \nafe\ and \nife .  The origin of such stars 
and a possible explanation of a tight correlation between \nife\ and \nafe\
are discussed in Paper I. 

The scatter in \cfe\ and \ofe\
is probably related to variations in the ratio between volatile and refractory
elements at a given age. Such variations were revealed by 
\citet{melendez09}, who compared \xfe\ for the Sun as a function of
the elemental condensation temperature, $T_{\rm c}$, in a solar
composition gas \citep{lodders03} 
with the corresponding trend for eleven solar twins having ages similar
to the age of the Sun. The Sun was found to have a higher
volatile-to-refractory ratio than most of the twins, which may
be due to sequestration of refractory elements 
in the terrestrial planets of the solar system \citep{melendez09}
or dust-gas segregation in star-forming clouds \citep{onehag14, gaidos15}.
Other examples of variations in the volatile-to-refractory ratio
(or the $\xh - \Tc $ slope) between stars of similar age are
presented by \citet{spina16b} in a high-precision study of 
14 solar twins. Differences of the $\xh - \Tc $ slope are also seen
between twin stars in binary systems \citep[e.g.][]{tuccimaia14, teske16a, teske16b, saffe17},
which is particular interesting, because dust-gas segregation before planet
formation is likely to affect the abundances of the two components
in the same way making it more likely that the observed abundance differences are
due to planet-star interactions.

\begin{figure}
\resizebox{\hsize}{!}{\includegraphics{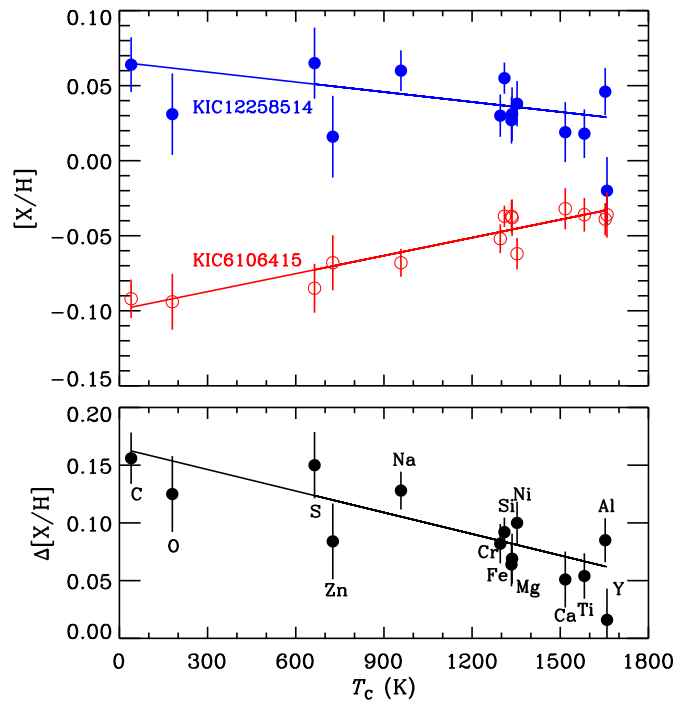}}
\caption{Comparison of the \xh -\Tc\ relations for KIC\,6106415 and
KIC\,12258514. The lower panel shows the difference in \xh\ between
KIC\,12258514 and KIC\,6106415. The lines show weighted least-squares fits to the data.}
\label{fig:122-610}
\end{figure}

In our sample of {\em Kepler} stars, \object{KIC\,6106415} and
\object{KIC\,12258514} provide an interesting case. The two stars have
about the same age as the Sun ($4.8 \pm 0.6$\,Gyr and $4.5 \pm 0.8$\,Gyr, respectively),
but as seen from Fig. \ref{fig:122-610}, the \xh -\Tc\ slopes
are very different: $+4.0 \, \pm 0.6 \cdot  10^{-5}$dex\,K$^{-1}$
for \object{KIC\,6106415} and $-2.2 \, \pm 1.1\cdot  10^{-5}$dex\,K$^{-1}$
for \object{KIC\,12258514}. Such variations make a large contribution to the
scatter in the age relations for \cfe\ and  \ofe , and also add to the
scatter in age relations of other volatile elements (Na, S, and Zn),
whereas abundance ratios between refractory elements (e.g. \mgfe )
are not significantly affected.

\begin{figure}
\resizebox{\hsize}{!}{\includegraphics{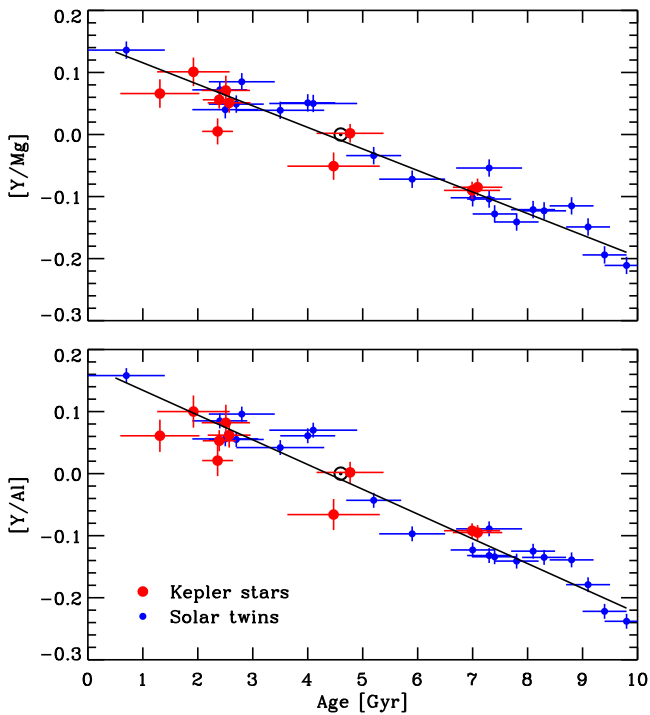}}
\caption{\ymg\ and \yal\ versus stellar age  for {\em Kepler} LEGACY stars
and solar twins 
including three $\alpha$-enhanced with ages between 9 and 10\,Gyr.
The lines show the linear fits corresponding to Eqs. (1) and (2).}
\label{fig:ymgal-age}
\end{figure}

A particularly strong age dependence is found for the ratios between
the $s$-process element yttrium and the $\alpha$-capture elements
magnesium and aluminium. As seen from Fig. \ref{fig:ymgal-age},
the {\em Kepler} stars confirm the age trends of \ymg\ and \yal\ previously found for
solar twins \citep{nissen16}. Maximum likelihood fits
to all stars (excluding the Sun) with
errors in both coordinates taken into account give
\begin{eqnarray}
\ymg = 0.150 \, (\pm 0.007) - 0.0347 \, (\pm 0.0012) \,\,Age \,{\rm [Gyr]}, 
\end{eqnarray}
and
\begin{eqnarray}
\yal = 0.174 \, (\pm 0.008) - 0.0400 \, (\pm 0.0012) \,\, Age \, {\rm [Gyr]}. 
\end{eqnarray}
In both cases, the reduced chi-square of the fit is 1.2, and the slope coefficients
are similar to those determined for solar twins alone
\citep{nissen16, spina16a}.

The strong increase of Y/Mg and Y/Al with decreasing stellar age, about
0.3\,dex over the lifetime of the Galactic disk, may be explained by 
production of yttrium via the slow neutron-capture process in low-mass
(1\,-\,4\, $M_{\rm Sun}$) asymptotic-giant-branch (AGB) stars
\citep[e.g.][]{travaglio04, karakas16}.
As time goes on, an increasing number of low-mass stars reach the AGB phase,
whereas the number of type II SNe producing Mg and Al declines.
It should, however, be emphasized that the tight relations
shown in Fig. \ref{fig:ymgal-age} were obtained for stars selected
to have metallicities in the range $-0.15 < \feh < +0.15$. \citet{feltzing17}
have shown that \ymg\ decreases with \feh\ at a given age.
This makes sense, because at lower metallicity there are fewer Fe atoms
per neutron to act as seeds for the $s$-process.  Hence, the use of \ymg\
(or \yal ) as a clock to measure stellar ages requires a calibration
that includes a metallicity term.

\begin{figure}
\resizebox{\hsize}{!}{\includegraphics{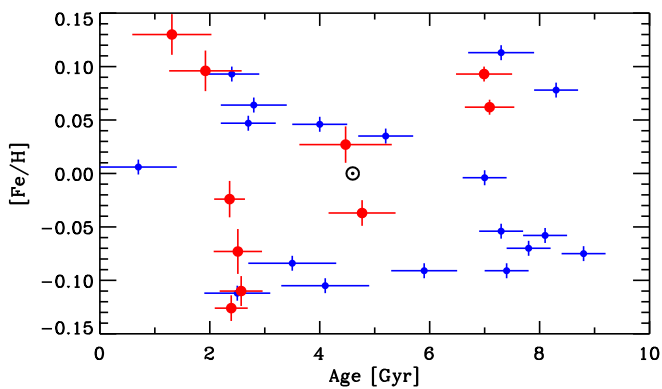}}
\caption{\feh\ versus stellar age  for {\em Kepler} LEGACY stars
(large red filled circles) and solar twins (small blue filled circles).}
\label{fig:feh-age}
\end{figure}

The tight relations between stellar age and abundance
ratios such as Mg/Fe and Y/Mg stand out in contrast to the lack of correlation 
between \feh\ and age (see Fig. \ref{fig:feh-age}).
The scatter in the age-metallicity relation for stars in the 
solar neighbourhood\,\footnote{According to Gaia DR1 parallaxes 
\citep{gaia16a, gaia16b} all {\em Kepler} stars
included in this paper have distances less than 140\,pc.}
is often explained as due to radial mixing of stars in a Galactic disk 
with a gradient in \feh ; stars presently in the solar neighbourhood
were born at different Galactocentric distances ($R_{\rm G}$).
However, in order to explain the low scatter in Mg/Fe and Y/Mg at a given age, 
the gradients in these abundance ratios should then be vanishingly small
at all times, which requires that
the ratios between the number of  Type II SNe , Type Ia SNe, and  AGB stars 
have evolved in the same way independent of $R_{\rm G}$. This corresponds to
a weak or non-existent inside-out formation history of the Galactic disk
\citep{haywood15}. 

\subsection{16\,Cyg\,A and B}
Among the {\em Kepler} stars included in this paper, the two components of the \object{16\,Cyg}
binary system are of particular interest. The very detailed set of
oscillation frequencies for these relatively bright stars has made it possible
to derive  precise values of masses, radii and ages \citep{metcalfe15}, helium abundances
\citep{verma14}, and rotation periods \citep{davies15}. Furthermore, a comparison 
of the seismic radii with those determined from interferometric angular diameters
\citep{white13} provides an important test of asteroseismology \citep[see][]{silva-aguirre17}.

The chemical compositions of 16\,Cyg\,A and B are also very interesting.
\citet{cochran97} discovered that \object{16\,Cyg\,B} has a Jupiter-sized planet
with a mass $M > 1.5 M_{\rm Jup}$
in an eccentric orbit with a period of 800 days, whereas no planets
have been detected around \object{16\,Cyg\,A}. Hence, a comparison of abundances
of the two stars has the potential of revealing possible effects of
planet formation or planet accretion on stellar surface composition
as discussed  by \citet{laws01}, who in a high-precision 
spectroscopic analysis found a small metallicity difference, 
$\Delta \feh$ (A -- B) = $0.025 \pm 0.009$, between 16\,Cyg\,A and B. 
This was supported by \citet{ramirez11}, who found systematic differences of the 
abundances of the two components corresponding to a mean difference 
of $\langle \Delta \xh \rangle$ (A -- B) = $0.041 \pm 0.007$ for 22 elements.
However, in a similar study by \citet{schuler11}, the mean abundance difference for 
15 elements was found to be only $0.003 \pm 0.015$\,dex.

In order to clarify the controversy about the chemical compositions of 
16\,Cyg\,A and B, \citet{tuccimaia14} made a new high-precision
spectroscopic analysis using CFHT/ESPaDOnS spectra with $R = 81\,000$ and $S/N = 700$.
From a strictly differential analysis of the two stars, they obtained
$\Delta \feh$ (A -- B) = $0.047 \pm 0.005$ supporting the abundance difference
found by \citet{ramirez11}. Furthermore, they found the abundance difference between 
A and B to increase with elemental condensation temperature
for the refractory elements, and interpreted this as a signature of
sequestration of refractory elements in the core of the giant planet 
around \object{16\,Cyg\,B}.

\begin{table}
\caption[ ]{Comparison of atmospheric parameters and \feh\ for 16\,Cyg\,A and B
between  \citet{tuccimaia14} and this paper.}

\label{table:16Cyg}
\centering
\setlength{\tabcolsep}{0.30cm}
\begin{tabular}{lcc}
\noalign{\smallskip}
\hline\hline
\noalign{\smallskip}
       &  16\,Cyg\,A  & 16\,Cyg\,B          \\
\noalign{\smallskip}
\hline
\noalign{\smallskip}
 \teff \,\tablefootmark{a}  & $5830 \pm 11$\,K & $5751 \pm 11$\,K  \\
 \teff \,\tablefootmark{b}  & $5816 \pm 10$\,K & $5763 \pm 10$\,K  \\
\noalign{\smallskip}
 \logg \,\tablefootmark{a} & $4.300 \pm 0.02$ & $4.350 \pm 0.02$  \\
 \logg \,\tablefootmark{b} & $4.291 \pm 0.01$ & $4.356 \pm 0.01$ \\
\noalign{\smallskip}
 \feh \,\tablefootmark{a} & $0.101 \pm 0.008$ & $0.054 \pm 0.008$ \\
 \feh \,\tablefootmark{b} & $0.093 \pm 0.007$ & $0.062 \pm 0.007$ \\
\noalign{\smallskip}
\hline
\end{tabular}
\tablefoot{
\tablefoottext{a}{\citet{tuccimaia14}.}
\tablefoottext{b}{This paper.}
}
\end{table}

The atmospheric parameters and abundances of 16\,Cyg\,A and B derived in this
paper from HARPS-N spectra with $R = 115\,000$ and $S/N \simeq 800$ 
support the results of \citet{tuccimaia14}. As seen from Table \ref{table:16Cyg},
the derived values of \teff , \logg , and \feh\ agree within the 
estimated one-sigma errors.  In particular, there is a remarkable agreement
between the gravities considering that \citet{tuccimaia14} used the ionization
equilibrium of Fe to obtain \logg , whereas our values are derived from seismic
data. Our A--B difference in \teff\ (53\,K) is somewhat smaller
than that of Tucci Maia et al. (79\,K), which leads to a smaller difference in \feh ,
but our difference $\Delta \feh$ (A -- B) = $0.031 \pm 0.010$ is still
significant at the three-sigma level.
There is also good agreement between the abundances of 13 elements in
common between the two works. The \xh -values agree within about one-sigma
of the combined errors, except in the case of oxygen, for which
the \oh\ values derived by Tucci Maia et al. are about 0.05\,dex
larger than our values.   

\begin{figure}
\resizebox{\hsize}{!}{\includegraphics{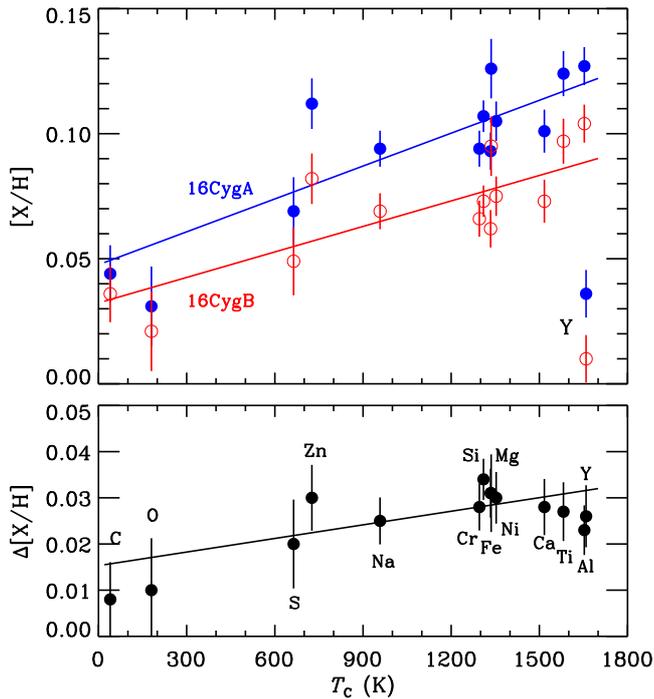}}
\caption{\xh\ values of 16\,Cyg\,A and B versus elemental condensation temperature.
The lower panel shows the difference in \xh\ between A and B.
The lines show weighted least-squares fits to the data (excluding yttrium).}
\label{fig:xh-Tc}
\end{figure}

As seen from Fig. \ref{fig:xh-Tc}, the \xh\ values of 16\,Cyg\,A and B
show an increasing trend as a function of the condensation temperature 
for the elements, as also found by \citet{tuccimaia14}. 
The scatter around the fitted lines is,
however, larger than expected from the estimated errors. Large deviations
are seen for Mg, Al, Zn, and in particular Y, 
that is the elements for which \xfe\ has the steepest dependence
on stellar age. Hence, the scatter is probably due to the age difference between
the Sun and the \object{16 Cyg} system. Interestingly, the difference in \xh\
between \object{16\,Cyg\,A} and \object{16\,Cyg\,B} (lower panel of Fig. \ref{fig:xh-Tc}) has
a much tighter correlation with condensation temperature, and increase
with increasing $T_{\rm c}$ at a slope of  
$+0.98 \, \pm 0.35 \cdot  10^{-5}$dex\,K$^{-1}$. In comparison,
\citet{tuccimaia14} found a $\Delta \xh$\,(A-B) $T_{\rm c}$-slope of 
$+1.88 \, \pm 0.79 \cdot  10^{-5}$dex\,K$^{-1}$ 
for refractory elements with $T_{\rm c} > 900$\,K. Our slope is smaller
but refers to the whole $T_{\rm c}$-range\,\footnote{Based on new 
spectra of 16\,Cyg\,A and B with $R \sim 160\,000$ and $S/N \sim 700$, 
\citet{tuccimaia17} obtain a $\Delta \xh$\,(A-B) $T_{\rm c}$-slope of
$+1.48 \, \pm 0.25 \cdot  10^{-5}$dex\,K$^{-1}$ for the whole $T_{\rm c}$-range
in better agreement with our result.}.
  
While the abundance differences between 16\,Cyg\,A and B
discussed above are small, there is a much larger difference for lithium
\citep[e.g.][]{king97, ramirez11}. As shown by \citet{carlos16}, \object{16\,Cyg\,B}
lie on the $A$(Li)-age relation for the large majority of solar twins,
but \object{16\,Cyg\,A} lie about 0.6\,dex above this relation. According to
\citet{deal15}, the small mass difference between A and B 
cannot explain the difference in $A$(Li). As discussed by 
\citet{melendez17}, the difference could, however, be due to a recent 
accretion by \object{16\,Cyg\,A} of a
Jupiter-sized planet in which the original amount of Li had been retained,
whereas 16\,Cyg\,B has not been exposed to such accretion.
This scenario may also explain the enhancement of 
heavier elements in \object{16\,Cyg\,A} relative to \object{16\,Cyg\,B}.
More detailed studies are, however, needed; \citet{deal15} argue that
fingering convection induced by accretion of heavy elements
will transport Li down to nuclear destruction layers and dilute
the pollution by heavy elements.

\section{Conclusions and outlook} 
\label{conclusions}
With the aim of testing trends of abundance ratios versus stellar age
as previously revealed for solar twins, we have used
HARPS-N spectra with $S/N\simgt250$ to derive high-precision
abundances for ten {\em Kepler} LEGACY stars having ages
determined from asteroseismology \citep{silva-aguirre17}. 
The seismic data have also been used to
determine precise values of stellar gravities, which allowed us
to derive effective temperatures from 
the ratio between equivalent widths of \FeI\ and \FeII\ lines.

The abundances were derived by an LTE analysis of spectral lines
using MARCS 1D model atmospheres. For the ranges of
atmospheric parameters covered by the stars 
($5750 < \teff  < 6350$\,K, $3.95 < \logg < 4.40$, and $-0.15 < \feh < +0.15$)
differential non-LTE corrections of abundances relative to those of the Sun
cannot be neglected.  Thanks to recent progress in non-LTE studies such
corrections are available for many of the elements included 
as discussed in Sect. \ref{non-lte}. The largest corrections of \xh\ for our sample
of stars are typically 0.02 to 0.04\,dex, which is of the same order of size
as the precision of the abundances derived. In addition to this, there
may also be significant differential 3D corrections, but  
these are only available in the case of oxygen \citep{amarsi16a}. Hence,  
3D non-LTE studies for other elements would be highly interesting to 
further improve the accuracy of the abundance determinations.
Effects of stellar activity \citep{flores16, reddy17} 
on abundance determinations also deserve further studies. 

We also note that atomic diffusion of elements
(gravitational settling and radiative acceleration) changes stellar
surface abundances as a function of time. According to the 
stellar evolution models of \citet{dotter17}, which include 
turbulent mixing and convective overshoot, \feh\ of
solar-type stars with $\teff < 6000$\,K decreases by approximately 0.10\,dex 
as the age increases from zero to 10\,Gyr, whereas \mgfe\ increases
by $\sim\!0.01$\,dex. Hence, diffusion cannot be neglected if we want
accurate information on Galactic chemical evolution or abundances at stellar birth.
Further studies of the change of stellar surface abundances as a function of age 
is much needed especially for $\teff > 6000$\,K, where the effect
of diffusion depends critically on whether one includes radiative
acceleration or not \citep[][Fig. 9]{dotter17}.

The abundances of the {\em Kepler} stars  confirm the trends between
\xfe\ and stellar age previously found for solar twins (see Fig. \ref{fig:xfe-age}).
For Mg, Al, Si, Ca, Ti, Cr, Zn, and Y,
the scatter in \xfe\ around fitted linear relations in an age interval
from 1 to 9\,Gyr is of the same order of size as the errors of the
\xfe\ abundance ratios, that is 0.01 - 0.02\,dex for solar twins 
and 0.02 - 0.03\,dex for {\em Kepler} stars. This small scatter 
stands out in contrast to a scatter of about 0.1\,dex in \feh\ 
(see Fig. \ref{fig:feh-age}). Often the scatter in the age-metallicity
relation for the solar neighbourhood is explained by radial mixing in a 
Galactic disk formed inside-out causing a gradient in \feh\ 
\citep[e.g.][]{minchev13, minchev14}, but it remains to be seen
if such models can explain the small scatter of for example \mgfe\ and \ymg .
As an alternative explanation 
of the scatter in the age-metallicity relation, \citet{edvardsson93}
suggest infall of metal-poor 
gas clouds onto the Galactic disk followed by stars formation before mixing
evens out the inhomogeneities in \feh.
If the clouds are sufficiently poor in metals, that is
\feh\ and $\mgh < -1$, infall coupled with star formation could cause a
scatter of 0.1\,dex in \feh\ (and \mgh ), but less than 0.01\,dex in
\mgfe .

The slopes of the \xfe -age relations (see Table \ref{table:fits}) 
provide new information on
the relative contribution of high- and low-mass supernovae to the nucleosynthesis
of elements. Whereas the slope is positive or near zero for most elements,
the $s$-process element yttrium stands out by having a strong negative 
abundance correlation with age. Over the lifetime of the Galactic disk, 
\ymg\ increases by about 0.3\,dex, which is probably due to an increasing
contribution of yttrium from low-mass AGB stars relative to the contribution
of magnesium from Type II SNe. This has led to the suggestion that \ymg\
may be used as a sensitive clock to find ages of stars
\citep{tuccimaia16, nissen16}. It should, however, be noted that 
the tight \ymg -age relation found in these papers and in the present
work refers to stars with metallicities close to solar. \citet{feltzing17}
have shown that the \ymg -age relation depends on metallicity, so  a  \feh\
term is needed for a proper calibration. Furthermore, the stars used
to discover the \ymg -age relation belong to the solar neighbourhood. 
A recent determination
of \ymg\ in K giants in four solar metallicity clusters with ages from 
1 to 6\,Gyr and distances up to 3\,kpc by  \citet{slumstrup17} suggests that
the \ymg -age relation is also valid outside the
solar neighbourhood, but this should be verified for a larger sample of distant stars.

While the scatter in the \xfe -age relations are small  for most elements, 
the scatter in \cfe\ and \ofe\ cannot be explained by observational
errors in abundances and ages. Thus, we found two stars 
(\object{KIC\,6106415} and \object{KIC\,12258514})
with about the same age and metallicity as the Sun, but with a difference 
$\Delta \cfe  = 0.092 \pm 0.028$ and a similar but less significant difference in \ofe .
Furthermore, the two stars have different slopes of \xh\ versus 
elemental condensation temperature \Tc\ (see Fig. \ref{fig:122-610}). 
We also found the two components in the binary star 16\,Cyg to have a
a small but significant difference in their \xh \,-\,\Tc\ slopes. These results 
point to variations in the volatile-to-refractory ratio in addition
to changes caused by chemical evolution, for example due to planet-star interactions.  
This is a complication in Galactic archaeology when using abundance
ratios for associating stars with a certain star-forming region
\citep[chemical tagging,][]{freeman02}. Abundance ratios of elements
with similar condensation temperatures, such as Mg/Fe and Y/Al, seem 
better suited for chemical tagging than abundance ratios between
elements with widely different values of \Tc , such as C/Fe and O/Fe,
but further studies of this problem are needed.

\begin{acknowledgements}
We thank Marcelo Tucci Maia for a constructive referee report and
Anish Amarsi for helpful comments on the discussion of non-LTE and 3D effects.
Based on observations made with the Italian Telescopio Nazionale Galileo
(TNG) operated on the Island of La Palma by the Fundaci\'{o}n Galileo Galilei of the INAF
(Instituto Nazionale di Astrofisica) at the Spanish Observatorio del Roque de los Muchachos 
of the Instituto de Astrofisica Canarias.
Funding for the Stellar Astrophysics Centre is provided by The
Danish National Research Foundation (Grant agreement no.: DNRF106).
This work has made use of data from the European Space Agency (ESA)
mission {\it Gaia} (\url{https://www.cosmos.esa.int/gaia}), processed by
the {\it Gaia} Data Processing and Analysis Consortium (DPAC,
\url{https://www.cosmos.esa.int/web/gaia/dpac/consortium}). Funding
for the DPAC has been provided by national institutions, in particular
the institutions participating in the {\it Gaia} Multilateral Agreement.
\end{acknowledgements}

\bibliographystyle{aa}
\bibliography{nissen.AA31845}

\end{document}